# The importance of crowding in signaling, genetic, and metabolic networks

Pieter Rein ten Wolde and Andrew Mugler

FOM Institute AMOLF, Science Park 104, 1098 XG, Amsterdam, The Netherlands

It is now well established that the cell is a highly crowded environment. Yet, the effects of crowding on the dynamics of signaling pathways, gene regulation networks and metabolic networks are still largely unknown. Crowding can alter both molecular diffusion and the equilibria of biomolecular reactions. In this review, we first discuss how diffusion can affect biochemical networks. Diffusion of transcription factors can increase noise in gene expression, while diffusion of proteins between intracellular compartments or between cells can reduce concentration fluctuations. In push-pull networks diffusion can impede information transmission, while in multi-site protein modification networks diffusion can qualitatively change the macroscopic response of the system, such as the loss or emergence of bistability. Moreover, diffusion can directly change the metabolic flux. We describe how crowding affects diffusion, and thus how all these phenomena are influenced by crowding. Yet, a potentially more important effect of crowding on biochemical networks is mediated via the shift in the equilibria of bimolecular reactions, and we provide computational evidence that supports this idea. Finally, we discuss how the effects of crowding can be incorporated in models of biochemical networks.

## Introduction

Cellular function arises at the level of the network of interactions between proteins and DNA inside the cell. Whether a cell differentiates or proliferates, stays put or moves, is not determined by a single component or reaction, but rather by the collective dynamics of signaling pathways and gene regulation networks. These biochemical networks are often modeled as well-stirred chemical reactors. However, cellular networks are fundamentally different from well-stirred chemical reactors. The copy numbers of the components are often low, and, as a result, biochemical networks can be highly stochastic (1-5). Moreover, the interior of the cell is a highly crowded environment, with proteins and DNA occupying up to 20-30% of the cellular volume (6, 7). These crowded conditions strongly affect the chemical reactions and physical interactions that form the biochemical network. Yet, while the effects of crowding on the dynamics of individual molecules and reactions have been investigated extensively (8), how crowding affects the collective dynamics of biochemical networks and thereby cellular function has remained largely unexplored.

Crowding can affect biochemical networks in two principal ways. First, crowding shifts the equilibrium of association-dissociation reactions to the associated state. Molecules in their associated state take up less volume, thereby creating more space for the other molecules in the system. This increases the entropy of the system and leads to an effective attraction between the molecules which is often called a depletion interaction or an excluded volume interaction (9); this depletion interaction effectively increases the binding affinity between two molecules. Second, crowding changes the diffusive motion of the molecules. At long length and timescales, diffusion appears normal, which means that the mean squared displacement increases linearly with time, just as in a dilute system without crowding agents; however, the slope of this linear relation, which gives the effective diffusion constant, can be orders of magnitude lower, because the crowders obstruct molecular motion (8, 10, 11). This lower diffusion constant decreases the association rate of diffusion-limited bimolecular reactions. On short length and timescales, on the other hand, crowding can lead to anomalous diffusion, or, more precisely, subdiffusion, which means that the mean-squared displacement grows sublinearly with time (12-15). This implies that it takes longer for initial positions to be forgotten, and that molecules that have come together, stay together longer.

While the effects of crowding on binding affinities and the diffusion of individual molecules have been studied in considerable detail both theoretically and experimentally, the role of crowding in the behavior of biochemical networks has only begun to be addressed. Yet, a number of studies in recent years have predicted that diffusion can have dramatic and often counterintuitive effects on the dynamics of signaling pathways, gene regulation networks and metabolic networks. In this review, we will summarize these studies. We hope to explain the phenomena using simple, intuitive arguments, and we will discuss how the crowded environment of the cell is expected to affect them. These studies also shed light on the question how crowding should be described in models of biochemical networks. Do the crowding agents have to be described explicitly or can the effect of their presence be captured by renormalizing the association and dissociation rates, thus allowing for a simpler model? Finally, we address the question whether the dominant effect of crowding on the dynamics of biochemical networks is the change in the diffusion dynamics or the shift in the binding equilibria.

**Diffusion of transcription factors can increase noise in gene expression**

Experiments in recent years have vividly demonstrated that gene expression can be highly stochastic (1-4). In particular, both in prokaryotes and eukaryotes, genes are often expressed in bursts. The burst-like nature of gene expression can emerge from the fact that one mRNA transcript can give rise to more than one protein copy – translation then occurs in bursts (2). Alternatively, the production of the mRNA itself can occur in bursts, as has been observed for prokaryotes (16-19). This can happen when the promoter switches between an "inactive" and an "active" state at timescales that are slower than the timescale of transcription initiation when the promoter is in the active state: then, more than one mRNA transcript will be produced from the promoter before the promoter switches back to the inactive state. While detailed time-series of mRNA production is consistent with such a two-state switching model of the promoter (17), the origin of the slow switching timescales that are necessary for bursty transcription is still unknown. One hypothesis is that it is due to the diffusion of transcription factors.

Many genes are regulated by transcription factors. The transcription-factor concentrations can be as low as a nanoMolar, or even lower. At such low concentrations, the stochastic arrival of transcription factors at the promoter can be a major source of noise in gene expression. In recent years, several groups have begun to address this question by theory and computer simulations (20-23). These studies revealed that below a microMolar, the diffusion of transcription factors can be the dominant source of noise in gene expression (21). These modeling studies also provide an explanation for why diffusion can contribute to the noise in transcription (Fig. 1). When a transcription factor dissociates from the promoter, it has a high probability of rebinding the promoter instead of diffusing into the bulk of other molecules. As a consequence, a transcription factor will undergo many rounds of rebinding and dissociation before it diffuses away from the promoter and escapes into the bulk. These rounds of rebinding and dissociation increase the effective active and inactive times of the promoter, which means that the mRNA transcripts are more likely to be produced in bursts -- long periods in which multiple mRNA copies are produced, alternating with long periods in which hardly any mRNA copies are produced. Crowding will decrease the diffusion constant and thereby decrease the association rate of the transcription factor with the DNA. It will also increase the number of rebindings and thereby decrease the effective dissociation rate (15, 23-25). Both effects will slow down the dynamics of the promoter state, with slower switching between the active and the inactive state of the promoter. This will lead to more bursts in transcription. As such, crowding has been predicted to increase the noise in gene expression (21, 23).

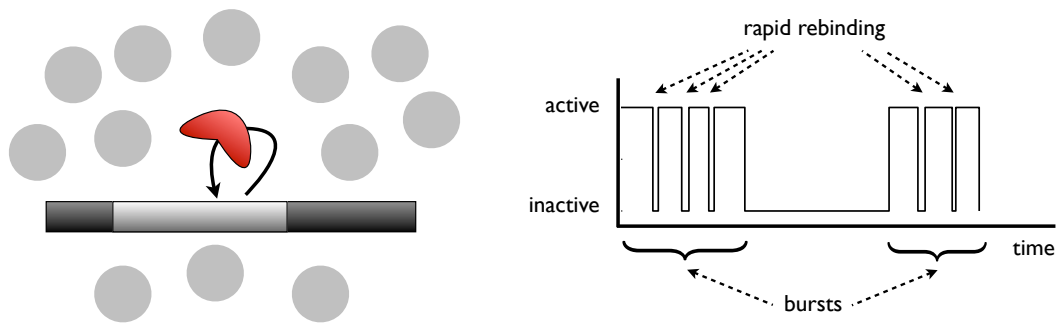

Figure 1. A transcription factor (red) diffusively interacting with the promoter region of DNA will undergo several rounds of dissociation and rapid rebinding before ultimately escaping to the bulk. As a result, gene expression will occur in bursts -- alternating periods of high and low transcription. Crowders (grey) enhance bursts by promoting rebinding and slowing diffusion.

**Diffusion between compartments can reduce protein concentration fluctuations**

Diffusion not only underlies bursts in gene expression, it can also act to "wash out" bursts across multiple compartments. Arguably the first hint of this effect was an experimental study on the precision of the *hunchback* (*hb*) expression domain in the developing embryo of the fruitfly Drosophila. Shortly after fertilization, the morphogen protein Bicoid (Bcd) forms an exponential concentration gradient along the anterior-posterior axis of the embryo, which provides positional information to the differentiating nuclei. One of the target genes of Bcd is *hb,* which is expressed in the anterior half of the embryo. The posterior boundary of the Hb expression domain is very sharp: by cell cycle 13, the position of the boundary varies only by about one nuclear spacing (26-28). This precision is higher than the best achievable precision for a time-averaging based readout mechanism of the Bcd gradient (27). Intriguingly, the study of Gregor et al. revealed that the Hb concentrations in neighboring nuclei exhibit spatial correlations, and the authors suggest that this implies a form of spatial averaging, which enhances the precision of the posterior Hb boundary (27). However, the mechanism for spatial averaging remained unclear.

To elucidate the mechanism of spatial averaging, Erdmann et al. (29) and independently Okabe-Oho et al. (30) performed a computational study of the Bcd-Hb system. Their analysis revealed a simple but robust mechanism for spatial averaging, which is based on the diffusion of Hb itself. The expression of Hb in the respective nuclei is, according to their analysis, produced in bursts. The diffusion of Hb between neighboring nuclei can, however, wash out these bursts in gene expression by rapidly distributing the protein copies that are produced during one burst in a nucleus over the neighboring nuclei (Fig. 2). Concomitantly, the fluctuations in the Hb concentration in a given nucleus depend not only on the expression of Hb in that nucleus, but, due to the diffusive exchange, also on the expression of Hb in the neighboring nuclei. Since Hb expression in the different nuclei is, to a good approximation, uncorrelated, this means that the concentration in a nucleus represents an average over the fluctuations in that nucleus and the neighboring nuclei. This is the mechanism of spatial averaging. Its strength increases with increasing diffusion constant, because the number of neighboring nuclei that contribute to the average concentration in a nucleus increases with the diffusion constant.

Interestingly, the computational study by Erdmann et al. also suggested that there exists an optimal diffusion constant that maximizes the precision of the boundary (29). On the one hand, a higher diffusion constant reduces the fluctuations in the Hb concentration in each nucleus via the mechanism of spatial averaging. On the other hand, a higher diffusion constant also lessens the steepness of the boundaries of the Hb expression domain. The tradeoff between these two effects yields an optimal diffusion constant that maximizes the sharpness or precision of the Hb expression domain.

The spatial averaging mechanism revealed by Erdmann et al. and Okabe-Oho et al. is very generic, and applies not only to any developmental system, but also to any biochemical network in general. For example, if a messenger protein is activated at one end of the cell and then has to diffuse to another place to activate another system, then the bursts in the activation of the messenger protein will be tempered by its diffusion. For this reason, it may be beneficial to spatially separate the input and output of the signaling pathway. Similarly, the diffusive exchange of signaling proteins between different intracellular compartments or between neighboring cells in a tissue will also tend to reduce the fluctuations in their concentration. A similar mechanism has also been implied in enhancing the fidelity of quorum sensing (31). Crowding will slow down diffusion and thus decrease the strength of spatial averaging.

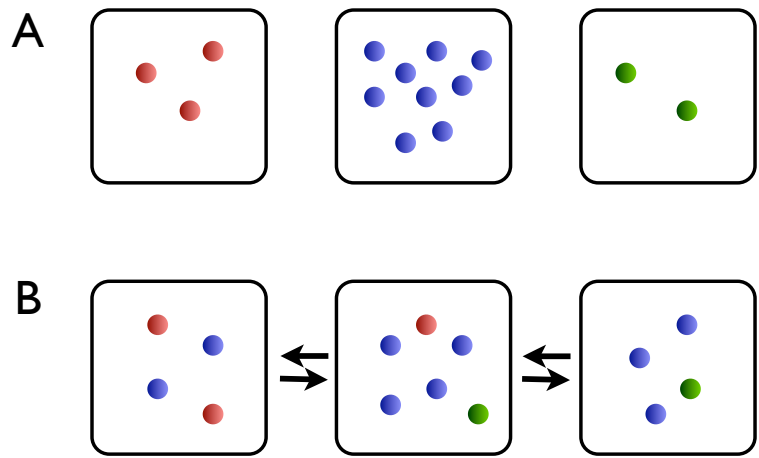

Figure 2. (A) In a developing Drosophila embryo, proteins in each cell nucleus are produced in bursts, meaning that some nuclei transiently have more proteins than others. (B) Diffusion of proteins between neighboring nuclei "washes out" these bursts, reducing transient fluctuations in protein number, a mechanism known as spatial averaging. Crowding will slow down diffusion and thus decrease the strength of spatial averaging. In both A and B, proteins are identical but are colored according to the compartment where they have been produced.

**Crowding can enhance information transmission by removing correlations**

Since the activation and expression of proteins is often irregular or bursty, the argument of spatial averaging may seem to suggest that diffusion of the activated or expressed protein is always beneficial. However, this is not necessarily the case, as the phenomenon of membrane protein clustering shows.

Membrane-bound proteins are often clustered into lipid domains or partitioned via cytoskeleton-induced corralling. Intriguingly, experiments have revealed that these clusters have a characteristic size of only a few molecules. For example, the GPI-anchored receptor CD59 is observed to form clusters of 3-9 molecules upon interaction with the cytoskeleton and lipid rafts (32, 33). Similarly, the well-studied membrane-bound GTPase Ras forms clusters of 6-8 molecules, which also depend on interactions with the cytoskeleton and rafts (34, 35). While several theoretical studies had predicted that clustering modulates signaling (36-39), the origin of the characteristic cluster size remained unknown.

To elucidate the origin of the characteristic cluster size, Mugler et al. performed a computational study of a common motif in membrane signaling. It consists of two layers of a push-pull network (40), in which the active component of the top layer activates the component of the bottom layer (Fig. 3). This motif can be found, for example, in CD59 and Ras membrane signaling. Stimulated CD59 receptors – the component of the top layer – induce the switching of several Src family kinases from an unphosphorylated to a phosphorylated state (32, 33) – the component of the bottom layer . Similarly, stimulated EGF receptors (top component) induce the switching of Ras proteins from an inactive GDP-loaded state to an active GTP-loaded state (38) (bottom component).

The authors found that spatially partitioning the components into domains or compartments not only linearizes the input-output relation, but also reduces the noise in the total concentration of the output. This is perhaps surprising, because partitioning the components reduces the size of the reaction volume (the partition), which tends to increase the fluctuations. Indeed, the fluctuations per partition do increase, as compared to a scenario without compartments and all molecules well mixed. Yet, the fluctuations in the output summed over all partitions decrease. The reason is that partitioning isolates molecules, as a result of which the molecules in the different compartments are activated independently. This reduces correlations in the switching behavior between their active and inactive state, and it is this removal of correlations that lowers the output noise (Fig. 3).

While these results show that it can be beneficial to spatially separate molecules, it does not explain the characteristic cluster size. The characteristic size suggests that there exists a tradeoff between two competing effects. But what is the effect that competes with the beneficial effect of noise reduction? The work of Mugler et al. suggests it is signal attenuation: making the partitions finer not only reduces noise, but also increases the probability that in a partition no signal can be propagated because only molecules of the top layer or only molecules of the bottom layer are present in that partition. This tradeoff between noise reduction (and linearization) on the one hand and signal attenuation on the other hand, naturally leads to an optimal cluster size that maximizes signaling fidelity and information transmission (40). Interestingly, the predicted optimal cluster size is in very good agreement with the characteristic cluster sizes that have been found experimentally for the Ras system and the CD59 system.

Spatial partitioning of membrane signaling proteins can be induced by the underlying cytoskeletal network or by membrane domains such as lipid rafts. However, even in a spatially homogeneous system, the finite speed of diffusion can lead to an effective partitioning of the molecules, which can enhance information transmission. The effect of partitioning is also not limited to membrane signaling. Partitioning may also be induced by scaffold proteins or by large macromolecular complexes in the cytoplasm. Each scaffold or macromolecular protein complex would act as an independent reaction compartment, and the exchange between these compartments would be mediated by rare dissociation events followed by rapid cytoplasmic diffusion and fast association with other complexes in the cytoplasm.

Crowding may play a key role in all these scenarios. Crowding can enhance the residence time of molecules on scaffolds and in macromolecular complexes, thereby allowing for the effective spatial isolation of molecules on the timescale of biochemical signaling. Even in the absence of scaffolds or macromolecular complexes, crowding is expected to be important. Crowding slows down diffusion, which leads to a finer effective partitioning of the signaling molecules. This may allow for the more effective removal of correlations between the signaling molecules, enhancing the fidelity of signaling.

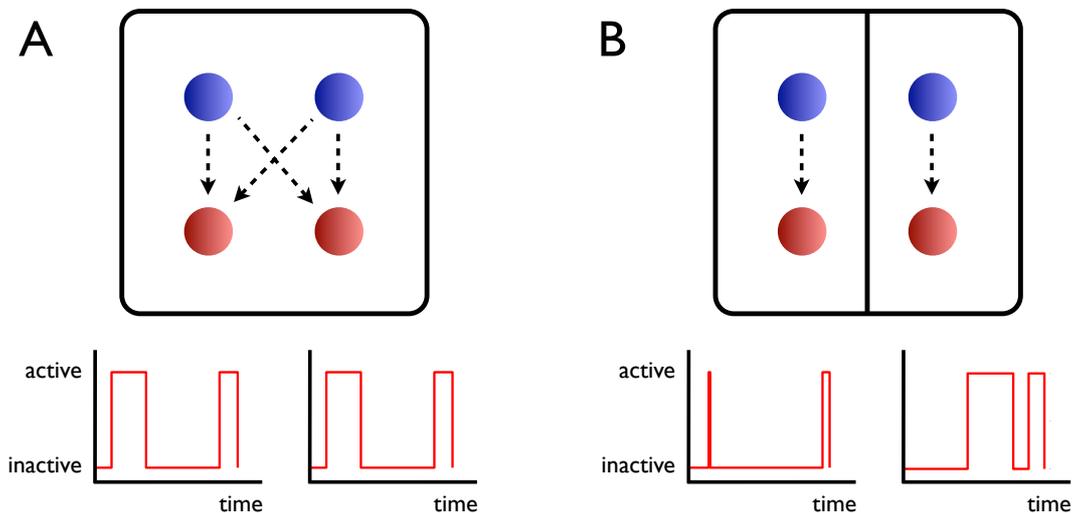

Figure 3. (A) In a typical signaling module, an upstream component (blue) activates a downstream component (red). Since both downstream molecules respond to both upstream molecules, the activation states of the two downstream molecules will be correlated with each other in time. (B) Spatial partitioning, which can be induced or enhanced by crowding, isolates molecules, thereby reducing these correlations and lowering the noise in the average activation across all downstream molecules.

**Crowding can promote membrane rebinding, which can enhance downstream signal propagation**

Clustering of membrane-bound proteins can not only enhance signal transmission in the membrane, but also change signal propagation from the membrane into the cytoplasm. A canonical motif in cell signaling is one in which a signaling protein in the cytoplasm is chemically modified by two antagonistic enzymes, with one enzyme, e.g. a kinase, residing at the membrane and the other, the phosphatase, in the cytoplasm. In some systems, such as the bacterial chemotaxis system, the cytoplasmic protein needs to be modified at only one site to become active, while in other systems, such as the Ras-Raf-Mek-Erk system and other MAPK signaling pathways (41), the cytoplasmic protein needs to be modified at multiple sites for full activation. It turns out that the effect of protein clustering on signal propagation strongly depends on the activation scenario (37).

When the cytoplasmic protein has to be modified at only a single site, then clustering of the activating, membrane-bound enzyme reduces the response, taken to be the concentration of the activated protein in the cytoplasm. This is a bulk effect: the clustering of the enzyme molecules at the membrane reduces the effective target size for the substrate molecules in the bulk (Fig. 4). This slows down the diffusive search, reducing the response. However, when the cytoplasmic molecules have to be modified at more than one site to become active, then clustering of the membrane-bound enzyme molecules can enhance the response. This is a rebinding effect: a cytoplasmic molecule that has been modified at it its first site, may rapidly rebind a signaling molecule in the cluster before it diffuses into the bulk where it is most likely deactivated by the deactivating enzyme. In a multi-site protein modification network, clustering thus allows for rapid rebindings, which may overcome the detrimental effect of the reduced target size. Crowding will, because of subdiffusion, promote rebinding (24, 25). These observations suggest that crowding can enhance the output of multi-site protein modification networks.

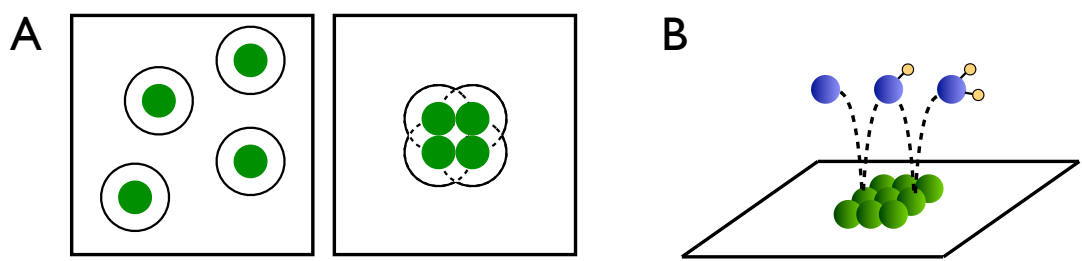

Figure 4. (A) Clustering of enzyme molecules reduces their effective target size for substrate molecules in the bulk, which decreases substrate activation. (B) On the other hand, clustering also enhances rapid substrate rebinding, which increases substrate activation if multi-site modification is required. Crowding further promotes rebinding, and thus may allow the latter effect to overcome the former.

**Crowding can qualitatively change the response of biochemical networks**

Cellular systems that exhibit macroscopic concentration gradients, such as those occurring during embryonic development or macroscopic spatio-temporal oscillations as in the Min system of E. coli, are typically treated as reaction-diffusion systems. In contrast, cellular systems that appear to be homogeneous at the cellular scale -- i.e. they do not exhibit concentration gradients across the cell -- are typically modeled as well-stirred chemical reactors, with the Gillespie algorithm being the most powerful numerical technique to simulate these models (42). In these well-stirred descriptions, it is assumed that at each instant in time the molecules are randomly distributed in space. This assumption seems reasonable when there are no macroscopic concentration gradients at the cellular scale. While it is commonly appreciated that many cellular systems exhibit spatial structures such as scaffold proteins that tend to localize near the membrane, in many signaling pathways and gene regulatory networks the components do, in fact, have a uniform concentration profile on average. It then only seems natural to simulate these systems as well-stirred chemical reactors.

However, this approach ignores the fact that even in a system that does not exhibit macroscopic concentration gradients, spatio-temporal correlations between the molecules may develop. These correlations mean that while each component individually may on average be uniformly distributed in space, some molecules tend to be close to other molecules, perhaps only transiently. Recent experimental and computational studies have shown that such spatio-temporal correlations at the molecular scale can drastically change the macroscopic behavior of the system at the macroscopic cellular scale (43-47). This phenomenon can arise in multi-site protein modification networks.

Multi-site protein modification networks are omnipresent in living cells. Well-known examples are the Kai system (48), the cyclin-dependent kinase inhibitor Sic1 (49), the Nuclear Factor of Activated T-cells system (50), and the CaM kinase II system (51). Arguably the best-known example is the MAPK pathway. MAPK pathways are involved in cell differentiation, cell proliferation, and apoptosis (41), and they exhibit very rich dynamics. It has been predicted mathematically and shown experimentally that they can generate an ultrasensitive response (52-54) and exhibit bistability via positive feedback (55). It has also been predicted that they can generate oscillations (56, 57), amplify weak but attenuate strong signals (58), and give rise to bistability due to enzyme sequestration (59, 60).

A typical MAPK pathway consists of three layers where in each layer a kinase activates the kinase of the next layer. Importantly, full activation of the kinase requires that it becomes doubly phosphorylated. This is regulated via a dual phosphorylation cycle in which the upstream kinase and a phosphatase control the modification state of the two sites of the kinase (the substrate) in an antagonistic fashion. It had long been believed that the response of the system strongly depends on whether the enzymes act in a processive or in a distributive manner (52-54). In a distributive mechanism, the enzyme has to release the substrate after it has modified the first site, before it can rebind and modify the second site. In contrast, in a processive mechanism, the enzyme remains bound to the substrate in between the modification of the two sites. Whereas a processive mechanism requires only a single enzyme-substrate encounter for the modification of both sites, a distributive mechanism requires at least two enzyme-substrate encounters. This may seem a small difference, but it can have marked implications. While a processive scheme gives rise to a hyperbolic response, a distributive scheme may give rise to a sigmoidal, ultrasensitive response (52-54). Moreover, a distributive scheme can, in contrast to a processive scheme, give rise to bistability, meaning that the distribution of phosphoforms becomes bimodal (59) . However, the studies that predicted all these behaviors assumed that the system can be described as a well-stirred chemical reactor. They thus ignored spatio-temporal correlations, which, as it turns out, can drastically change the behavior of the system.

To assess the importance of spatio-temporal correlations, Takahashi et al. performed a computational study of one layer of a MAPK cascade (43). In their model, the enzymes operate according to a distributive scheme. They found, however, that spatio-temporal correlations can reduce the response of the system to that of a system in which the enzymes act according to a processive scheme: ultrasensitivity and bistability may be lost in regimes where the well-stirred model predicts that they should be present. The origin of this effect is rapid enzyme-substrate rebinding (Fig. 5). After a kinase has phosphorylated a substrate molecule at its first site, the two molecules will dissociate. The kinase then has to exchange ADP for ATP before it can bind the substrate molecule again to phosphorylate the second site. During this refractory time, the kinase and the substrate molecule will diffuse away from each other. When the refractory time is long and/or the diffusion constant is high, the kinase and the substrate molecule will have diffused away from each other into the bulk by the time the kinase becomes active again. In this case, the second site of the substrate molecule will most likely be phosphorylated by another kinase molecule. This is the scenario that is captured by well-stirred chemical models. However, when the refractory time is short and/or the diffusion constant is low, then the kinase and the substrate molecule will probably still be in close physical proximity to each other when the kinase becomes active again (Fig. 5). The kinase and the substrate molecule can then rebind. Importantly, this means that both sites are phosphorylated by the same kinase molecule. In this case, the response of the system becomes that of a processive scheme and ultrasensitivity and bistability are lost. The distinction between a processive scheme and a distributive one is thus not whether the enzyme and the substrate molecule remain physically connected to each other in between the modification of the two sites, but rather whether the two sites are phosphorylated by the same enzyme molecule -- the enzyme and the substrate molecule may diffuse away from each other and back in between the modification steps.

The loss of bistability can be understood at a mechanistic level. In a well-stirred system, bistability arises due to enzyme sequestration. Imagine, for example, that the system is in the state in which most substrate molecules are unphosphorylated; in this state, the kinase molecules are mostly bound to substrate molecules, in the process of phosphorylating them, while the phosphatase molecules are freely diffusing (Fig. 5). The question then is what will happen to a substrate molecule that has just been phosphorylated by a kinase at its first site: will it bind a kinase that will phosphorylate it further or a phosphatase that will dephosphorylate it, driving it back to its original state? In the well-stirred model, the spatial distribution of the molecules is not important, and one only needs to consider the number of kinase and phosphatase molecules that are available for binding. Clearly, in the well-stirred model, the singly phosphorylated substrate molecule will most likely bind a phosphatase, because the phosphatase molecules are free while the kinase molecules are sequestered by the other substrate molecules. In the spatially-resolved model, however, it may be more likely that the singly phosphorylated molecule binds a kinase molecule – indeed the kinase molecule that just phosphorylated it and is still nearby – even though there are many more phosphatase molecules that it could bind to.

These results show that spatio-temporal correlations due to rapid rebindings can drastically change the macroscopic behavior of a multi-site protein modification network. Crowding can enhance rebindings (24, 25), which means that crowding can change the response of these multi-site protein modification networks (43-45, 47). Recently, this prediction was confirmed experimentally by Aoki et al. (44, 47). They studied the Raf-Mek-Erk MAPK pathway in Hela cells and found the response to be processive. To elucidate the effect of crowding, they also studied the same pathway in a test tube, both in the presence and in the absence of the crowding agent polyethylene glycol (PEG). In the absence of PEG, the response was observed to be distributive. However, in the presence of PEG, the response became processive, precisely as predicted by Takahashi et al. (43). These beautiful experiments unambiguously demonstrate that crowding can have a dramatic effect on the response of biochemical networks.

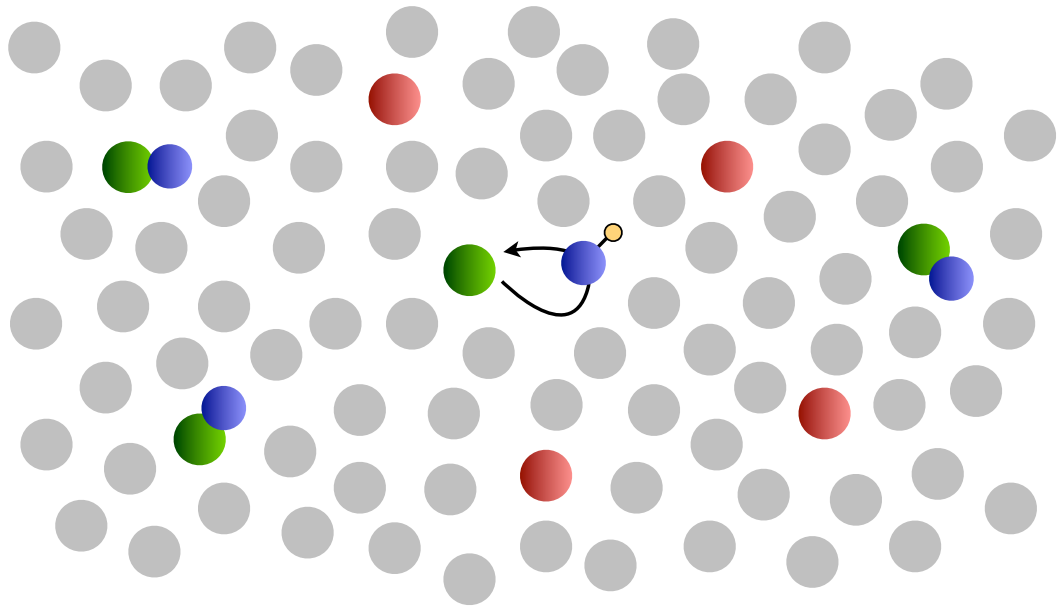

Figure 5. A substrate molecule (blue) has just been phosphorylated by a kinase (green) in a system where most phosphatases (red) are free and most kinases are sequestered by other substrate molecules. In a well-stirred model, the substrate molecule is more likely to bind a phosphatase because there are more free phosphatases than free kinases. In contrast, in a spatially-resolved model, the substrate molecule is more likely to rapidly rebind the kinase because it remains nearby. These rebindings can lead to the loss of bistability, i.e. the loss of bimodality in the distribution of substrate phosphoforms. Crowders (grey) promote rebinding and, as such, they can qualitatively change the macroscopic response of the system.

**Diffusion can affect the metabolic flux**

Crowding is expected to not only affect cell signaling, but also the metabolic flux. It has long been thought that the spatial colocalization of metabolic enzymes might increase the metabolic flux, minimize the pool of unwanted intermediate products, and reduce transient time scales (61, 62). New experimental techniques have now made it possible to also study these effects quantitatively by experiment. Advances in single-molecule techniques enable the investigation (63) and manipulation (64) of the dynamics of individual enzyme molecules. In parallel, molecules can now be positioned with nanometer precision in artificial systems, along one-dimensional channels, on two-dimensional surfaces (65), and with DNA origami even in three dimensions (66, 67). These experimental developments now allow the detailed characterization of how metabolic flux depends on the spatial distribution of the enzymes and how this is affected by crowding.

Recently, Buchner et al. began to address this question using a simple theoretical one-dimensional model of a metabolic network (68). In this model, a metabolite enters the system at one end; the metabolite then diffuses through the system until it either leaves the other end or is converted by an enzyme into the desired product. The simplicity of the model permitted a detailed analysis, which revealed that the effect of enzyme clustering can be highly nontrivial. In particular, they found that the optimal arrangement of the enzyme molecules depends on a parameter that characterizes the probability that the enzyme reacts with the metabolite versus the probability that the metabolite is lost by diffusion. When this parameter is high, meaning that the catalytic rate of the enzyme is high, then product formation is maximized by uniformly distributing the enzyme molecules over the system. However, when this parameter is low, then it is advantageous to cluster the enzyme molecules near the origin where the metabolite enters the system. The authors explain these observations via the concept of enzyme exposure, which quantifies how much the metabolite molecules are exposed to the enzyme molecules (Fig. 6). Most molecules only stay a relatively short time in the system, because they quickly leave the system at the other end; this means that most metabolite molecules cover the entire space of the system. Hence, when the catalytic rate is high, and metabolites can be quickly converted into product, it becomes beneficial to spread the enzyme molecules over the system. In contrast, when the catalytic rate is low, metabolite and enzyme interact a large number of times before a reaction takes place. Metabolite molecules that quickly leave the system at the other end will thus not be converted; only those metabolite molecules that happen to stay in the system for a long time have a significant chance to react with the enzyme. The metabolite molecules that happen to stay long in the system are those that by chance remain close to the origin. It then becomes beneficial to cluster the enzymes close to the origin. The authors also found that for intermediate values of the control parameter – the catalytic rate over the diffusion speed – the optimal enzyme profile is an arrangement in which a fraction of the enzyme molecules is clustered near the origin, while the other fraction is approximately uniformly distributed over the system.

Related work shows that similar effects also occur in systems with different geometries, as well as higher dimensions (Buchner et al., private communication). What is of particular interest from the perspective of crowding is that recurrence or residence time is a key concept – this was indeed precisely the reason why the clustered configuration is optimal when the catalytic rate is low. Crowding can enhance the residence time, which means that crowding is expected to also affect how the flux in metabolic networks depends on the spatial distribution of the enzymes.

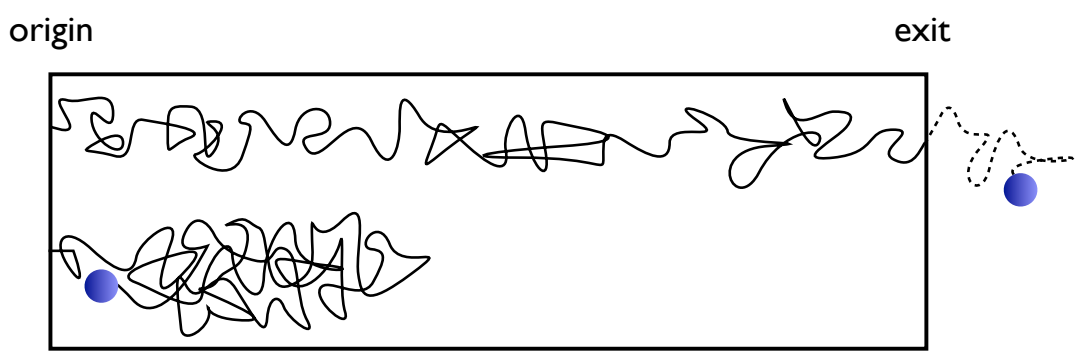

Figure 6. Metabolite molecules (blue) enter the left end of a one-dimensional channel and diffuse until either exiting the right end or being converted by an enzyme into a product. Most metabolite molecules reach the exit (top trajectory). Therefore when conversion is fast, product formation is optimized by a uniform enzyme distribution. When conversion is slow, however, only those metabolites that happen to stay long in the system have any chance of being converted. The molecules that stay long in the system are those that happen to spend much time near the origin (bottom trajectory). Hence, when conversion is slow, it becomes optimal to put the enzymes near the origin. Diffusive residence times in this and other geometries are significantly enhanced by crowding.

## How to model biochemical networks in the presence of crowding

The above results show that diffusion and crowding can significantly change the behavior of biochemical networks. This suggests that biochemical networks cannot be treated as well-stirred chemical reactors and that spatially resolved models are necessary. However, while in recent years efficient numerical techniques have been developed to simulate spatially-resolved models (43, 60, 69-72), simulating these models remains computationally challenging. The question arises whether the effects of diffusion cannot be captured by renormalizing the association and dissociation rates and then use these effective rates in a well-stirred model, which could then be simulated using the powerful Gillespie algorithm (42).

To answer this question we need to inspect the collective dynamics of the network on short and long time scales and compare this to the dynamics of the diffusive motion of the network components on these timescales. With regard to diffusion, recent theoretical (73) and experimental (10, 74) work shows that while the diffusion constant is decreased by the presence of the crowders, the molecules still move by normal diffusion on long length- and timescales. Elf et al. showed that in E. coli the Lac repressor moves by normal diffusion on timescales longer than 10 ms (10) and Bancaud et al. showed that also in eukaryotic nuclei the proteins move by normal diffusion on timescale longer than 20 ms (74). At these long length and timescales, the molecules have lost memory of where they came from, and to a good approximation they have randomly distributed themselves over the system, just as in a well-stirred chemical reactor. Consequently, we expect that the distribution of association times is exponential at long times even in the crowded environment of the cell (21, 43). On short length and timescales, however, crowding will enhance the probability of rebinding (15, 23, 24). Such rapid rebindings lead to an algebraic distribution of association times on short timescales (21, 43); moreover, the probability of rebinding does not depend on the concentration. In contrast, in a well-stirred model, all association reactions depend on the concentration and are exponentially distributed on all timescales. The question then becomes whether the algebraic diffusion dynamics on these short time scales, originating from rebindings, can qualitatively change the behavior of the network.

As we have seen above, rebindings can qualitatively change the macroscopic behavior of the system in multi-site protein modification networks (43, 45, 46). Here, a single rebinding event can cause irreversible modification of a species, which then takes the system from one state to another, effectively putting the system onto a different dynamical pathway (75). Also for the metabolic network problem discussed above, the recurrence or return of trajectories to the origin (the source), which underlies the benefit of clustering the enzyme molecules, cannot be captured in a well-stirred model (68). These systems have to be treated as reaction-diffusion systems, even though they may not exhibit concentration gradients at the cellular scale.

However, in many biochemical networks, the state of the system is the same before and after rebinding. Moreover, the timescale of rebinding is typically much faster than the timescale that governs the system dynamics. If both conditions are met, it is possible to treat a group of rebinding events as a single effective event and to renormalize the association and dissociation rates by the number of rebindings (21, 75). An example is the binding of a transcription factor (TF) to the DNA: the TF-DNA complex is identical before and after a rebinding event and the rebinding time is much shorter than the protein life time, making it possible to integrate out the rebindings. Indeed, a spatially-resolved model of gene expression that explicitly describes the diffusive motion of a TF in space gave essentially identical results to a well-stirred model with renormalized TF-DNA association and dissociation rates (21). Hence, when rebindings do not put the system onto a different dynamical pathway, then the effect of diffusion and crowding may be captured in the association and dissociation rates in a zero-dimensional model, which can be simulated much more efficiently than a spatially resolved model.

**The effect of crowding: diffusion or entropy?**

The previous sections discussed how crowding could influence biochemical networks by introducing spatio-temporal correlations in the dynamics or by simply changing the magnitude of the diffusion constant. However, it is well known that crowding also shifts the equilibrium of association-dissociation reactions toward the associated state. This is largely because the associated state occupies less volume than the dissociated molecules. In the associated state, more volume is therefore available for the surrounding crowder molecules, thereby increasing the total entropy of the system (9). Importantly, most if not all, biochemical networks involve bimolecular reactions. This means that the shift in equilibria can have a pronounced effect on the behavior of biochemical networks. The question is whether this effect outweighs the effect of diffusion.

To address this question, Morelli et al. developed a simple scheme to model the effects of crowding on biochemical networks (76). It is based on the observation discussed above that even in the presence of crowding, the distribution of association and dissociation is expected to remain exponential on the relevant time scales of the network. This observation makes it possible to use a well-stirred model and capture the effect of crowding on both diffusion and the equilibrium shift by appropriately scaling the association and dissociation rates. They used this approach in combination with kinetic Monte Carlo simulations to analyze the effect of crowding on a constitutively expressed gene, a repressed gene, and a model for the bacteriophage λ switch. They found that the effects of crowding are mainly caused by the shift of association-dissociation equilibria rather than the slowing down of protein diffusion. While slower diffusion of the regulatory proteins increases the noise in gene expression (21), it does not significantly change the mean expression level, in contrast to the shift in the equilibrium of the binding of the RNA polymerase (RNAP) and the regulatory proteins to the promoter. The results on the constitutively expressed gene and the self-repressed gene are therefore not very surprising. The results on the phage switch λ are more surprising, because previous work has shown that the stability of the switch is very sensitive to the noise in gene expression (77, 78), which is increased by the slow down in diffusion. However, it turns out that the behavior of the switch is much more affected by the shift in the equilibria of the binding of the RNAP and the regulatory proteins to the promoter. The study also revealed that the result of these equilibrium shifts can be highly counterintuitive, because of their competing effects on the flipping behavior of the switch. For example, the authors found that the switching rate depends non-monotonically on the concentration of crowder. This observation is potentially much more generic: crowding will affect the equilibria of all bimolecular reactions in any biochemical network, but how this plays out in the collective dynamics of the network is difficult to predict by intuition.

**Outlook**

While the effect of crowding on the diffusion of individual molecules has been characterized extensively in the past decades, its importance for the dynamics of biochemical networks remains a wide open question. We hope that this review shows that crowding is expected to significantly alter the dynamics of signal transduction pathways, gene regulation networks and metabolic networks. As such, we hope that this review stimulates further research in this direction. A particularly promising approach is to reconstitute the biochemical network of interest in the test tube and to elucidate the importance of crowding by systematically varying the concentration of a crowding agent, such as poly-ethylene glycol (PEG) (44). Such in vitro studies have provided tremendous insight in the diffusion dynamics of individual molecules, and we expect that they will be similarly fruitful in unraveling the role of crowding in biochemical networks. Finally, in this review we have mostly focused on cellular systems that are homogeneous at the cellular scale. Yet, in many signaling pathways cellular concentration gradients exist because proteins are activated at one position but deactivated at another (79, 80); the range of these gradients is set by the diffusion constant over the deactivation rate, which means that crowding will also change the dynamics of these networks. Moreover, systems that exhibit spatio-temporal oscillations, such as the Min system (81), are also sensitive to the diffusion dynamics of the components and hence to crowding (82). Indeed, we expect that many interesting crowding-induced phenomena will be uncovered in the coming years.

## Acknowledgements

This work is part of the research program of Stichting voor Fundamenteel Onderzoek der Materie, which is financially supported by Nederlandse Organisatie voor Wetenschappelijk Onderzoek.

1. Elowitz MB (2002) Stochastic Gene Expression in a Single Cell. Science 297:1183–1186.

2. Ozbudak EM, Thattai M, Kurtser I, Grossman AD (2002) Regulation of noise in the expression of a single gene. Nature.

3. Paulsson J (2004) Summing up the noise in gene networks. Nat Cell Biol 427:415–418.

4. Yu J (2006) Probing Gene Expression in Live Cells, One Protein Molecule at a Time. Science 311:1600–1603.

5. Taniguchi Y et al. (2010) Quantifying E. coli Proteome and Transcriptome with Single-Molecule Sensitivity in Single Cells. Science 329:533–538.

6. Zimmerman SB, Trach SO (1991) Estimation of macromolecule concentrations and excluded volume effects for the cytoplasm of Escherichia coli. Journal of Molecular Biology 222:599–620.

7. Ellis RJ (2001) Macromolecular crowding: an important but neglected aspect of the intracellular environment. Current Opinion in Structural Biology 11:114–119.

8. Muramatsu N, Minton AP (1988) Tracer diffusion of globular proteins in concentrated protein solutions. P Natl Acad Sci Usa 85:2984–2988.

9. Asakura S, Oosawa F (1954) On interaction between two bodies immersed in a solution of macromolecules. J Chem Phys.

10. Elf J, Li GW, Xie XS (2007) Probing transcription factor dynamics at the single-molecule level in a living cell. Science 316:1191.

11. Bancaud AEL et al. (2009) Molecular crowding affects diffusion and binding of nuclear proteins in heterochromatin and reveals the fractal organization of chromatin. The EMBO Journal 28:3785–3798.

12. Weiss M, Elsner M, Kartberg F, Nilsson T (2004) Anomalous Subdiffusion Is a Measure for Cytoplasmic Crowding in Living Cells. Biophysical Journal 87:3518–3524.

13. Banks DS, Fradin C (2005) Anomalous Diffusion of Proteins Due to Molecular Crowding. Biophysical Journal 89:2960–2971.

14. Golding I, Cox E (2006) Physical Nature of Bacterial Cytoplasm. Phys Rev Lett 96:098102.

15. Bénichou O, Chevalier C, Klafter J, Meyer B, Voituriez R (2010) Geometry-controlled kinetics. Nature Chemistry 2:472-477.

16. Golding I, Paulsson J, Zawilski SM, Cox EC (2005) Real-time kinetics of gene activity in individual bacteria. Cell.

17. So L-H et al. (2011) General properties of transcriptional time series in Escherichia coli. Nat Genet 43:554–560.

18. Chubb JR, Trcek T, Shenoy SM, Singer RH (2006) Transcriptional Pulsing of a Developmental Gene. Current Biology 16:1018–1025.

19. Raj A, Peskin CS, Tranchina D, Vargas DY, Tyagi S (2006) Stochastic mRNA synthesis in mammalian cells. Plos Biol.

20. Bialek W, Setayeshgar S (2005) Physical limits to biochemical signaling. P Natl Acad Sci Usa 102:10040–10045.

21. van Zon JS, Morelli MJ, Ten Wolde PR (2006) Diffusion of transcription factors can drastically enhance the noise in gene expression. Biophysical Journal 91:4350–4367.

22. Tkačik G, Bialek W (2009) Diffusion, dimensionality, and noise in transcriptional regulation. Phys Rev E 79:051901.

23. Meyer B, Bénichou O, Kafri Y, Voituriez R (2012) Geometry-Induced Bursting Dynamics in Gene Expression. Biophysical Journal 102:2186–2191.

24. Lomholt MA, Zaid IM, Metzler R (2007) Subdiffusion and weak ergodicity breaking in the presence of a reactive boundary. Phys Rev Lett 98:200603.

25. Zaid IM, Lomholt MA, Metzler R (2009) How Subdiffusion Changes the Kinetics of Binding to a Surface. Biophysical Journal 97:710–721.

26. Houchmandzadeh B, Wieschaus E, Leibler S (2002) Establishment of developmental precision and proportions in the early Drosophila embryo. Nature 415:798–802.

27. Gregor T, Tank DW, Wieschaus EF, Bialek W (2007) Probing the Limits to Positional Information. Cell 130:153–164.

28. He F et al. (2008) Probing Intrinsic Properties of a Robust Morphogen Gradient in Drosophila. Developmental Cell 15:558–567.

29. Erdmann T, Howard M, Ten Wolde PR (2009) Role of Spatial Averaging in the Precision of Gene Expression Patterns. Phys Rev Lett 103:258101.

30. Okabe-Oho Y, Murakami H, Oho S (2009) Stable, precise, and reproducible patterning of bicoid and hunchback molecules in the early Drosophila embryo. … computational biology.

31. Weber M, Buceta J (2011) Noise regulation by quorum sensing in low mRNA copy number systems. Bmc Syst Biol 5:11.

32. Suzuki K, Fujiwara TK, Edidin M (2007) Dynamic recruitment of phospholipase Cγ at transiently immobilized GPI-anchored receptor clusters induces IP3–Ca2+ signaling: single-molecule tracking study 2. The Journal of cell ….

33. Suzuki KGN et al. (2007) GPI-anchored receptor clusters transiently recruit Lyn and G for temporary cluster immobilization and Lyn activation: single-molecule tracking study 1. J Cell Biol 177:717–730.

34. Prior IA (2003) Direct visualization of Ras proteins in spatially distinct cell surface microdomains. J Cell Biol 160:165–170.

35. Plowman SJ, Muncke C, Parton RG, Hancock JF (2005) H-ras, K-ras, and inner plasma membrane raft proteins operate in nanoclusters with differential dependence on the actin cytoskeleton. Proc Natl Acad Sci U S A 102:15500.

36. Kalay Z, Fujiwara TK, Kusumi A (2012) Confining domains lead to reaction bursts: reaction kinetics in the plasma membrane. PLoS ONE.

37. Mugler A, Bailey AG, Takahashi K, Ten Wolde PR (2012) Membrane clustering and the role of rebinding in biochemical signaling. Biophysical Journal.

38. Tian T et al. (2007) Plasma membrane nanoswitches generate high-fidelity Ras signal transduction. Nat Cell Biol 9:905–U60.

39. Gurry T, Kahramanoğulları O, Endres RG (2009) Biophysical mechanism for Ras-nanocluster formation and signaling in plasma membrane. PLoS ONE.

40. Mugler A, Tostevin F, Ten Wolde PR (2013) Spatial partitioning improves the reliability of biochemical signaling. Proc Natl Acad Sci U S A 110:5927–5932.

41. Chang L, Karin M (2001) Mammalian MAP kinase signalling cascades. Nature 410:37–40.

42. Gillespie DT (1977) Exact stochastic simulation of coupled chemical reactions. J Phys Chem 81:2340–2361.

43. Takahashi K, Tanase-Nicola S, Ten Wolde PR (2010) Spatio-temporal correlations can drastically change the response of a MAPK pathway. Proc Natl Acad Sci U S A 107:2473–2478.

44. Aokia K, Yamada M, Kunida K, Yasuda S, Matsuda M (2011) Processive phosphorylation of ERK MAP kinase in mammalian cells. Proc Natl Acad Sci U S A

108:12675–12680.

45. Hellmann M, Heermann DW, Weiss M (2012) Enhancing phosphorylation cascades by anomalous diffusion. EPL-Europhysics Letters.

46. Abel SM, Roose JP, Groves JT, Weiss A, Chakraborty AK (2012) The Membrane Environment Can Promote or Suppress Bistability in Cell Signaling Networks. J Phys Chem B 116:3630–3640.

47. Aoki K, Takahashi K, Kaizu K, Matsuda M (2013) A quantitative model of ERK MAP kinase phosphorylation in crowded media. Sci Rep 3:1541.

48. Zwicker D, Lubensky DK, Ten Wolde PR (2010) Robust circadian clocks from coupled protein-modification and transcription–translation cycles. P Natl Acad Sci Usa 107:22540–22545.

49. Nash P et al. (2001) Multisite phosphorylation of a CDK inhibitor sets a threshold for the onset of DNA replication. Nature 414:514–521.

50. Crabtree GR, Olson EN (2002) NFAT signaling: choreographing the social lives of cells. Cell 109 Suppl:S67–79.

51. Miller P, Zhabotinsky AM, Lisman JE, Wang X-J (2005) The Stability of a Stochastic CaMKII Switch: Dependence on the Number of Enzyme Molecules and Protein Turnover. Plos Biol 3:e107.

52. Huang C, Ferrell JE (1996) Ultrasensitivity in the mitogen-activated protein kinase cascade. P Natl Acad Sci Usa 93:10078–10083.

53. Ferrell JE (1996) Tripping the switch fantastic: how a protein kinase cascade can convert graded inputs into switch-like outputs. Trends in Biochemical Sciences 21:460–466.

54. Ferrell JE, Bhatt RR (1997) Mechanistic studies of the dual phosphorylation of mitogen-activated protein kinase. Journal of Biological Chemistry 272:19008–19016.

55. Ferrell JE, Machleder EM (1998) The biochemical basis of an all-or-none cell fate switch in Xenopus oocytes. Science 280:895–898.

56. Kholodenko BN (2000) Negative feedback and ultrasensitivity can bring about oscillations in the mitogen-activated protein kinase cascades. European Journal of Biochemistry.

57. Wang X, Hao N, Dohlman HG, Elston TC (2006) Bistability, Stochasticity, and Oscillations in the Mitogen-Activated Protein Kinase Cascade. Biophysical Journal 90:1961–1978.

58. Locasale JW, Shaw AS, Chakraborty AK (2007) Scaffold proteins confer diverse regulatory properties to protein kinase cascades. P Natl Acad Sci Usa 104:13307–13312.

59. Markevich NI (2004) Signaling switches and bistability arising from multisite phosphorylation in protein kinase cascades. J Cell Biol 164:353–359.

60. Elf J, Ehrenberg M (2004) Spontaneous separation of bi-stable biochemical systems into spatial domains of opposite phases. Syst Biol (Stevenage) 1:230–236.

61. Gaertner FH (1978) Unique catalytic properties of enzyme clusters. Trends in Biochemical Sciences.

62. Heinrich R, Schuster S, Holzhütter HG (1991) Mathematical analysis of enzymic reaction systems using optimization principles. Eur J Biochem 201:1–21.

63. Xie S (2001) Single-molecule approach to enzymology. Single Molecules.

64. Gumpp H, Puchner EM, Zimmermann JL, Gerland U (2009) Triggering enzymatic activity with force. Nano ….

65. Kufer SK, Puchner EM, Gumpp H, Liedl T, Gaub HE (2008) Single-Molecule Cut-and-Paste Surface Assembly. Science 319:594–596.

66. Müller J, Niemeyer CM (2008) DNA-directed assembly of artificial multienzyme complexes. Biochemical and Biophysical Research Communications 377:62–67.

67. Fu J, Liu M, Liu Y, Woodbury NW, Yan H (2012) Interenzyme Substrate Diffusion for an Enzyme Cascade Organized on Spatially Addressable DNA Nanostructures. J Am Chem Soc 134:5516–5519.

68. Buchner A, Tostevin F, Gerland U (2013) Clustering and Optimal Arrangement of Enzymes in Reaction-Diffusion Systems. Phys Rev Lett 110:208104.

69. Hattne J, Fange D, Elf J (2005) Stochastic reaction-diffusion simulation with MesoRD. Bioinformatics 21:2923–2924.

70. van Zon JS, Ten Wolde PR (2005) Simulating biochemical networks at the particle level and in time and space: Green's function reaction dynamics. Phys Rev Lett 94:128103.

71. van Zon JS, Ten Wolde PR (2005) Green's-function reaction dynamics: a particle-based approach for simulating biochemical networks in time and space. J Chem Phys 123:234910.

72. Erban R, Chapman SJ (2009) Stochastic modelling of reaction–diffusion processes: algorithms for bimolecular reactions. Phys Biol.

73. Kim JS, Yethiraj A (2009) Effect of macromolecular crowding on reaction rates: a computational and theoretical study. Biophysical Journal 96:1333–1340.

74. Bancaud A, Lavelle C, Huet S, Ellenberg J (2012) A fractal model for nuclear organization: current evidence and biological implications. Nucleic Acids Research 40:8783–8792.

75. Mugler A, Ten Wolde PR (2013) The Macroscopic Effects of Microscopic Heterogeneity in Cell Signaling. Advances in Chemical Physics.

76. Morelli MJ, Allen RJ, Ten Wolde PR (2011) Effects of Macromolecular Crowding on Genetic Networks. Biophysical Journal 101:2882–2891.

77. Warren PB, Ten Wolde PR (2005) Chemical Models of Genetic Toggle Switches †. J Phys Chem B 109:6812–6823.

78. Morelli MJ, Tănase-Nicola S, Allen RJ, Ten Wolde PR (2008) Reaction Coordinates for the Flipping of Genetic Switches. Biophysical Journal 94:3413–3423.

79. van Albada SB, Ten Wolde PR (2007) Enzyme Localization Can Drastically Affect Signal Amplification in Signal Transduction Pathways. PLoS Comput Biol 3:e195.

80. Kholodenko BN (2006) Cell-signalling dynamics in time and space. Nat Rev Mol Cell Biol 7:165–176.

81. de Boer PA, Crossley RE, Rothfield LI (1989) A division inhibitor and a topological specificity factor coded for by the minicell locus determine proper placement of the division septum in E. coli. Cell 56:641–649.

82. Weiss M (2003) Stabilizing Turing patterns with subdiffusion in systems with low particle numbers. Phys Rev E Stat Nonlin Soft Matter Phys 68:036213.